\documentstyle[12pt]{article}
\input epsf

\begin{document}
\date{\today}
\def\be{\begin{equation}}
\def\ee{\end{equation}}
\def\bea{\begin{eqnarray}}
\def\eea{\end{eqnarray}}
\def\a{\alpha}
\def\b{\beta}
\def\N{{\mathcal{N}}}
\def\t{{\triangle}}
\def\g{{\gamma}}
\def\l{{\lambda}}
\def\A{{\cal A}}


\newpage
\bigskip
\hskip 3.7in\vbox{\baselineskip12pt
\hbox{NSF-ITP-01-72}}

\bigskip\bigskip

\centerline{\large \bf  On the thermodynamic instability of  LST}

\bigskip\bigskip

\centerline{{\bf
Alex Buchel\footnote{buchel@itp.ucsb.edu}}}

\bigskip
\centerline{Institute for Theoretical Physics}
\centerline{University of California}
\centerline{Santa Barbara, CA\ \ 93106-4030, U.S.A.}

\begin{abstract}
\baselineskip=16pt
The high energy thermodynamics of Little String Theory (LST) is known
to be unstable. An unresolved question is whether the corresponding
instability in LST holographic dual is of stringy or supergravity
origin. We study UV thermodynamics of a large metric deformation of
the LST dual realized (in the extremal case) by type IIB fivebranes
wrapping a two-sphere of a resolved conifold, and demonstrate that the
resulting black hole has negative specific heat.  This explicitly
shows that the high energy thermodynamic instability of the LST
holographic dual is of the supergravity origin.

\end{abstract}
\newpage
\setcounter{footnote}{0}

\baselineskip=17pt

\section{Introduction}
Little string theory (LST) is  a  non-local  theory 
defined on the world volume of  NS5-branes in the limit of vanishing 
string coupling $g_{str}\to 0$ where the string scale 
$\a'$ is kept fixed \cite{brs,s9705} (for a review see \cite{a9911}).
This theory does not include dynamical gravity and in the IR 
flows to six-dimensional Yang-Mills theory. Nonetheless,
LST is quite different from a local field theory: 
it exhibits T-duality in toroidal compactifications and a 
Hagedorn density of states at very high energy --- both 
the intrinsic attributes
of the ``standard'' string theories. 
The latter property in particular implies that the 
statistical mechanics of LST breaks down at a finite 
temperature, known as the Hagedorn temperature.
The purpose of this paper is to better understand the
ultraviolet thermodynamics of the LST.

A classical thermodynamics of the LST on the world volume 
of flat NS5 branes can be easily deduced from its 
holographic dual, realized as a near horizon geometry 
of non-extremal NS5 branes \cite{ms9710}. It was shown in \cite{ms9710} that 
as  the number $N$ of the five-branes is large, $N\gg 1$, 
and the energy $\mu$ above 
extremality in string unites satisfies $\mu\gg N$, the 
background geometry is smooth with small curvatures everywhere; 
in addition, the string perturbation theory is also good 
as the dilaton is bounded from above by its value 
at the horizon, ${N\over \mu}$. Ignoring loop/stringy 
corrections, one finds that the Hawking temperature of the 
resulting black hole is independent of the energy $\mu$
and coincides precisely with the Hagedorn temperature $T_H$ of the 
LST 
\be
\b_H={1\over T_H}=2\pi\sqrt{N\a'}\,.
\label{th}
\ee
Since the can tune the energy and the temperature of the 
system independently, its equation of state (in the 
ultraviolet) is 
\be
S=\b_H E\,,
\label{seh}
\ee
which leads to an exponential growth of the density of 
states
\be
\rho(E)\sim e^{\b_H E}\,.
\label{ph}
\ee
One-loop corrections to the Hagedorn density of states of LST were 
studied in \cite{ho00,br00,ks00}. The finite energy corrections 
to the density of states (\ref{ph}) were  argued to be of the 
form 
\be
\rho(E)\sim E^{\a} e^{\b_H E}\biggl(1+O\left({1\over E}\right)\biggr)\,.
\label{pf}
\ee
The sign of $\a$ in (\ref{pf}) is of uttermost importance as it determines 
the stability of the thermal ensemble representing LST at 
high energy. The explicit calculation of \cite{ks00} indicated 
that $\alpha$ is negative implying the negative specific heat 
and thus the thermal instability of the system. 
According to authors of \cite{ks00} the instability in question is of stringy 
origin. It is represented by a (massless at tree level) 
string mode that winds once around the Euclidean time direction,
but is supposed to become tachyonic at one-loop level.     

An alternative explanation of the instability of the LST  
at high energy has been advocated in \cite{r0104}. 
Following the conjecture of \cite{gm09,gm11} that 
thermodynamic instabilities 
in field theories should correspond to classical instabilities of the 
dual spacetime geometry, Rangamani proposed \cite{r0104} that the 
supergravity dual to LST at Hagedorn 
temperature suffers from a Gregory-Laflamme (GL) \cite{gl9301,gl9404}  
like instability, thereby causing the thermal ensemble to be unstable. 
Though he did not manage to explicitly demonstrate that the metric 
fluctuations about the background of interest have indeed a zero 
frequency mode, that is capable of explaining the origin of the 
instability, Rangamani conjectured that the required mode 
is the one responsible for the GL instability in the 
near-extremal NS5-branes claimed in \cite{re0104}, that would survive 
the decoupling limit for the finite temperature LST.
 
In this paper we prove the proposal of \cite{r0104} and present 
yet another explicit example advocating the general philosophy 
of \cite{gm09,gm11}. We do not perform the stability 
analysis near the background representing 
finite temperature LST realized on  flat NS5 branes 
as suggested in \cite{r0104}. 
Rather, we study large metric deformations and 
show that the resulting black hole geometries 
have negative specific heat at the {\it classical} level.
Recall that in the flat case the temperature of the LST is (classically)
independent of the energy. By changing the 
theory in the IR (say wrapping NS5 branes on a 2-cycle, so that the  
theory is four dimensional macroscopically), while preserving 
the UV characteristics (having a 2-cycle of a finite size),
we generically expect that temperature should become energy dependent. 
This must definitely be the case, if the deformation 
induces a  phase transition in the IR, so that 
the deformed geometry is simply singular (and thus does not make sense) 
for sufficiently small energy. 
Then, by studying the classical thermodynamics of 
the deformed background at  high energy 
we should be able to deduce $\a$ in (\ref{pf}).
Specifically, we propose to study the stability 
issue of LST at high energy from the thermodynamics 
of large number of type IIB NS5 branes wrapping a two-cycle 
of a resolved conifold. The relevant extremal solution
was discussed in \cite{chv1,chv2,mn0008}. It has $\N=1$  SUSY 
in four dimensions and exhibits 
a naked singularity in the IR associated with the 
chiral symmetry breaking of the dual gauge theory at zero 
temperature. The nonextremal deformation of this background 
was constructed in \cite{bf0103}. It was argued there that the 
naked singularity (at the extremality) will be hidden beyond the black 
hole horizon for sufficiently large energy away from the extremality.

The paper is organized as follows. In the next section 
we describe the gravitational background representing 
a large number of type IIB NS5 branes wrapping a 2-cycle in the
resolved conifold geometry at finite temperature. We study 
 thermodynamics of this black hole solution, and 
show that at high energy the black hole has a negative specific 
heat. We analytically compute $\a$ in the  
density of states expression (\ref{pf}) for the discussed 
deformation. We briefly conclude in section 3. 
Technical details can be found in the appendix.

\section{High energy thermodynamics of $\N=1$ deformed LST}
In the previous section addressing the origin 
of thermodynamic instability of LST at high energy, 
we proposed to study  large metric deformation of its holographic dual 
realized  by wrapping NS5 branes on a two-sphere of a resolved conifold. 
The motivation for choosing this particular background 
came from the fact that the corresponding extremal 
solution had a naked singularity in the IR associated with 
the zero temperature chiral symmetry breaking phase transition 
of the dual gauge theory. As the result, the Hawking 
temperature of its nonextremal deformation should be energy 
dependent, and so the classical analysis should be enough 
to extract $\a$ in the ``corrected'' entropy-energy relation 
\be
S=\b_H E+\a \log E +O\left(1/E\right)\,.
\label{corr}
\ee  

We proceed by recalling the  gravitational 
background holographically dual to the $\N=1$
deformed LST at finite temperature,
originally constructed in \cite{bf0103}. From now on 
we set the string scale $\a'=1$. 
The ten dimensional string frame metric, NS-NS 3-form $H$, 
and the dilaton $\phi$
are given by\footnote{$\t,\ f(\t)$ used here are related 
to $c_1(r),\ h(r),\ a$ of \cite{bf0103} 
as follows  $\t\equiv \t_1(r)$, 
$f(\t)=c_1^8(r) \t_1(r) h(r)/A$, $N=a^2$.} 
\bea
ds^2_{str}&=&\t^2 dt^2-d\bar{x}^2-N\biggl[ f^2 (d[\ln\t])^2+
{h\over 4}\left(d\theta_1+\sin^2\theta_1 d\phi_1^2\right)\cr
&&+{1\over 4}\left(d\theta_2+\sin^2\theta_2 d\phi_2^2\right)
+{1\over 4}\left(d\psi+\sum_{i=1}^2 \cos\theta_i d\phi_i\right)^2\ \biggr]\,,\cr
\cr
H&=&{N\over 4}\biggl[\left(d\psi+\sum_{i=1}^2 \cos\theta_i d\phi_i
\right)\wedge\left(\sin\theta_1 d\theta_1\wedge d\phi_1-\sin\theta_2 
d\theta_2\wedge d\phi_2\right)\biggr]\,,\cr
\cr
e^{-2\phi}&=&C {f\over h \t }\,,
\label{m10}
\eea
where $N$ is the number of NS5 branes, $C$  is an integration 
constant related to the asymptotic  
string coupling,  and $\t$ is a 
new  radial coordinate defined in such a way that 
$\t=0_+$ determines the black hole horizon, and $\t\to 1_-$ is the 
ultraviolet asymptotic. 

From eq.~(5.56) of \cite{bf0103}, $f$ and $h$ satisfy 
second-order Toda-like system of differential equations 
\bea
&&\left[\ln{f}\right]''=4 f^2\left(1+{2\over h}-{1\over h^2}\right)\,,
\label{f}
\eea
\bea
&&\left[\ln h\right]''=4 f^2\left({2\over h}-{2\over h^2}\right)\,,
\label{h}
\eea 
supplemented by the first-order constraint 
\bea
0&=&\left[{f^2 \over h^2}\right]'\left[h^2\right]'+
{f^2 \over h^2}(h')^2+2 h^2 \left(\left[{f\over h}
\right]'\right)^2-2 f^2 h^2\cr
&&-8 {f^4 \over h^2}(h^2+2 h -1)\,.
\label{consi}
\eea
In eqs.~(\ref{f}),~(\ref{h}),~(\ref{consi}) 
the derivatives are with respect to 
\be
y\equiv \ln \t\,,\qquad y\in(-\infty,0)\,.
\label{y}
\ee
As explained in \cite{bf0103}, to have a regular Schwarzschild
horizon we must have 
\bea
&&f= e^y\left(f_0+\sum_{n=1}^{\infty}f_n e^{2y n}\right)\,,\cr
&&h= h_0+\sum_{n=1}^{\infty}h_n e^{2y n}\,,\qquad  {\rm as}\ y\to -\infty\,,
\label{hor}
\eea
for some positive constants\footnote{Only $h_0$ is an 
independent parameter,  $f_0$ should be fixed by the 
boundary conditions \cite{bf0103}.} $f_0,h_0$. In the UV region we 
have 
\bea
&&h\to +\infty\,,\qquad  {\rm as}\ y\to 0_{-}\,,\cr
&&f\to +\infty\,,\qquad 
{\rm as}\ h\to +\infty\,.
\label{ext}
\eea
Furthermore, it was argued in \cite{bf0103} that both 
the IR and the UV asymptotics are compatible provided  $h_0>1$,
where the inequality incorporates the physics of chiral symmetry 
restoration at finite temperature of the dual gauge theory.

We now study the thermodynamic properties of the black hole
(\ref{m10}) keeping the dilaton at the horizon $g_h$ fixed, for 
large $h_0\gg 1$. The first condition amounts to choosing 
\be
C={h_0\over g_h^2 f_0}\,,
\label{Cdef}
\ee
and as we will see insures (for $g_h\ll 1$) 
that the dilaton is everywhere small.
The second condition is the high energy limit of the thermodynamics.
We don't know how to solve (\ref{f}), (\ref{h}) in general. 
However, the $h_0\to \infty$ asymptotic of the solution can be 
extracted with some work. We find\footnote{See appendix for the 
details.}
\bea
h(y)&=&-2 y -2 \log[-\sinh y]+h_0+O(1)\,,\cr
\cr
f(y)&=&-{1\over 2\sinh y} \left(1-{1\over h(y)}+o\left({1\over h(y)}\right)
\right)\,.
\label{solution}
\eea
With (\ref{solution}) we see that the dilaton is indeed bounded 
by its value at the horizon, also 
\be
f(y) e^{-y}\bigg|_{y\to -\infty}\to f_0=1-1/h_0+o(1/h_0)\,. 
\label{fff}
\ee
From the  metric of (\ref{m10}) we can read off the Hawking temperature
by identifying the periodicity of its  Euclidean time direction 
with the inverse temperature  
\be
T^{-1}\equiv\b=2\pi N^{1/2} f_0=\b_H \left(1-{1\over h_0}+o\left({1\over 
h_0}\right)\right)\,.
\label{temp}
\ee 
Next, we compute the Bekenstein-Hawking entropy of the geometry 
(\ref{m10}). We find the 8-dimensional area of the event horizon $\A_8$
of the black hole to be
\be
\A_8=2 g_h^{-2} N^{5/2} \pi^3 h_0 V_3\,,
\label{area}
\ee
where $V_3$ is the 3-dimensional volume. The entropy of the 
black hole is then
\be
S={\A_8\over 4 G_N}={N^{5/2} h_0 V_3\over 16 \pi^3 g_h^4}\,.
\label{entropy}
\ee
From the ordinary statistical mechanics we know that the 
energy is $dE=T dS$, so from (\ref{temp}), (\ref{entropy})
we find the entropy-energy relation for our particular 
deformation of LST
\be
S(E)=\b_H E+\alpha\log E+o(\log E)\,,
\label{enen}
\ee 
with 
\be
\a=-{N^{5/2} V_3\over 16\pi^3 g_h^4}\,.
\label{compal}
\ee
Note that as in the flat LST case \cite{ks00}, $\alpha$ is 
and extensive quantity --- it is proportional to the noncompact 
 factor  $V_3$ of the fivebrane world-volume. 
Since $\a$ in (\ref{compal}) is negative, 
the black hole (\ref{m10}) has negative specific heat and 
is thus thermodynamically unstable.

\section{Conclusion}    
The purpose of this paper was to argue that the instability 
of the ultraviolet LST thermodynamics in its holographic 
dual  is  of a  supergravity origin, as suggested in \cite{r0104}.
To do so, we identified a large metric deformation of the 
gravity dual to LST where the (conjectured) threshold instability 
of the flat LST dual should be enhanced to classical instability 
of the nonextremal generalization of the deformed background.
We argued that this should happen if the deformation in the 
(extremal case) induces a phase transition in the IR, 
but does not affect the UV physics. In our example,
this large metric deformation is realized by a large number 
of NS5 branes wrapping a 2-cycle of a resolved conifold.  
By studying the UV thermodynamics of this background 
we explicitly showed that resulting black hole has a negative 
specific heat.
 
The nonextremal geometry of interest has small curvatures 
and small perturbative string loop corrections if both 
$1/N$ and $g_h$ are small. Thus our classical analysis 
is reliable.

\section*{Acknowledgments}
We wish to thank    Anton Kapustin, Igor Klebanov
and Arkady Tseytlin
for valuable discussions. This work  was supported in 
part by NSF grants PHY97-22022 and PHY99-07949.

\section*{Appendix}
In what follows we solve (\ref{f}), (\ref{h}) subject to the boundary 
conditions (\ref{hor}), (\ref{ext}) in the limit $h_0\to \infty$.
Note that from (\ref{f}), (\ref{h}) both 
$f$ and $h$ are rapidly increasing functions, so in the 
zero order approximation in the limit $h_0\to \infty$, we can set
$1/h=0$ in (\ref{f}). This equation has then a general solution 
\be
f\approx f^{(0)}=-{C_1\over 2 \sinh(C_1 y+C_2)}\,,
\label{solf0}
\ee
where $C_i$ are integration constants.
From the UV asymptotic we need $f\to\infty$ as $y\to 0_-$. 
This fixes $C_2=0$. Furthermore, the asymptotic at the horizon 
fixes $C_1=1$ so that  
\be
f^{(0)}=-{1\over 2 \sinh y}\,.
\label{ff0}
\ee
Note that  satisfying the asymptotics 
uniquely fixes $f_0$,  
\be
f_0^{(0)}=1\,.
\label{f00}
\ee 
In the leading order eq.~(\ref{h}) reads
\be
[\ln h^{(0)}]''={2\over h^{(0)} \sinh^2 y}\,.
\label{hh0}
\ee
Above equation is difficult to solve analytically. 
We claim however that in the limit $h_0\to \infty$
\be
[h^{(0)}]'' h^{(0)}\gg ([h^{(0)}]')^2\,,
\label{ineq}
\ee 
uniformly for all $y$. In approximation (\ref{ineq}), eq.~(\ref{hh0})
simplifies 
\be
[h^{(0)}]''={2\over  \sinh^2 y}\,.
\label{hsimp}
\ee
The general solution of (\ref{hsimp}) is 
\be
h^{(0)}=C_1 y-2 \log[-\sinh y]+C_2\,,
\label{h01}
\ee
where $C_i$ are integration constants. To satisfy 
asymptotic at the horizon we must choose $C_1=-2$ and $C_2=h_0-2 \ln 2$
so that 
\be
h^{(0)}=-2 y-2 \log[-\sinh y]+h_0-2\ln 2\,.
\label{hof}
\ee
We can now go back to (\ref{ineq}) to check the self-consistency 
of the approximation. 
Indeed, we find that ${[h^{(0)}]'' h^{(0)}\over  ([h^{(0)}]')^2}$
with $h^{(0)}$ given by (\ref{hof}) has a global minimum 
at $y=\left(-1/h_0+O(h_0^{-2})\right)$, with the value at the minimum 
$\left(h_0/2-\log h_0+ O(1)\right)\to \infty$ as $h_0\to \infty$.

To study the leading correction to the Hagedorn 
thermodynamics we actually need to know the first correction 
to (\ref{ff0}). We take the following ansatz for the leading 
correction
\be
f=-{1\over 2 \sinh y} \left(1+{\gamma\over h^{(0)}}+o({1\over h^{(0)}})
\right)\,,
\label{1st}
\ee
where $h^{(0)}$ is the zero order solution (\ref{hor}).
Substituting (\ref{1st}) into (\ref{f}), using eq.~(\ref{hsimp})
and the approximation (\ref{ineq}), we  find $\gamma=-1$. 
In a similar fashion we can compute the leading correction 
to $h^{(0)}$ and see that it is a fixed constant.  
From (\ref{1st}) it follows that 
\be
f_0=1-{1\over h_0}+o\left({1\over h_0}\right)\,.
\label{1stf0}
\ee

Since our conclusion about UV thermodynamic instability of LST 
hinges on the fact that $\g<0$, we also did numerical 
analysis to confirm (\ref{1stf0}). In the remaining of this
section we  describe  them and present the results. 
First, we rewrite (\ref{f}) and (\ref{h}) in terms of 
\bea
&&f_1(t)\equiv {e^y\over f(y)}\,,\cr
&&f_2(t)\equiv {1\over h(y)}\,,
\label{red}
\eea
where we introduced new variable 
\be
t\equiv e^{2 y},\qquad t\in [0,1]\,.
\label{tdef}
\ee
We find 
\bea
0&=&t f_1(t) [f_1(t)]''+f_1(t) [f_1(t)]'-t ([f_1(t)]')^2+1+2 f_2(t)
-f_2(t)^2\,,\cr
\cr
0&=&t f_1(t)^2 f_2(t) [f_2(t)]''+f_1(t)^2 f_2(t) [f_2(t)]'-t f_1(t)^2 
([f_2(t)]')^2\cr 
&&+2 f_2(t)^3-2 f_2(t)^4\,,
\label{eq}
\eea
where all the derivatives are with respect to $t$.
Near  $t=0$ we have power series expansion
\bea
f_1=\sum_{k=0}^{\infty} \a_k t^k\,,\qquad f_2=\sum_{k=0}^{\infty} 
\b_k t^k\,,
\label{t0}
\eea
with 
\bea
&&\a_0={1\over f_0},\qquad \a_1=-{f_0 (h_0^2+2 h_0-1)\over h_0^2},\qquad 
\a_2=\cdots\,,\cr
&&\b_0={1\over h_0},\qquad \b_1=-{2 f_0^2 (h_0-1)\over h_0^3},\qquad 
\b_2=\cdots\,,
\label{init}
\eea
while at $t\to 1_-$ we have boundary condition
\be
0<f_1(t)\ll f_2(t)\to 0\,.
\label{bound}
\ee

\begin{figure}
\centering
\epsfxsize=2.5in
\hspace*{0in}\vspace*{.2in}   
\epsffile{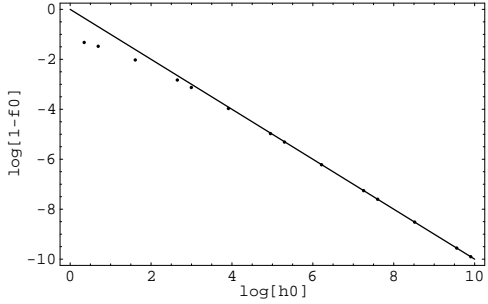}   
\caption{Numerical determination of $f_0(h_0)$.
Dots represent values of $\log[1-f_0(h_0)]$  evaluated  
as detailed in the appendix section.  
The slope is predicted to be $(-1)$ from (\ref{1stf0}).}
\end{figure}

In practice we integrated (\ref{eq}) from $t=\delta t=10^{-4}$
with initial conditions, determined by ${f_0,h_0}$, specified 
by the first six terms in the expansion (\ref{t0}). 
Notice that is if a set  $\{f_1(t,[f_0,h_0]), f_2(t,[f_0,h_0])\}$
is a solution to (\ref{eq}), then so does 
$\{f_1(t\l,[f_0/\sqrt{\l},h_0]), f_2(t\l,[f_0/\sqrt{\l},h_0])\}$ 
for arbitrary $\l$. The latter in particular implies 
that if $f_1(t=1,[f_0,h_0])=0$ (as it should be for the 
boundary condition (\ref{bound})), than 
$f_1(t=\l,[f_0/\sqrt{\l},h_0])=0$. This suggested a 
practical ``trick'' of fixing $f_0$ in terms of $h_0$ from 
the integration: for a fixed $h_0$ we take $f_0=1$ and find 
(via numerical integration) $t^*\equiv t^*(h_0)$ such that 
$f_1(t^*(h_0),[1,h_0])=0$; then, above considerations guarantee that 
$f_1(1,[\sqrt{t^*(h_0)},h_0])=0$. That is 
\be
f_0(h_0)=\sqrt{t^*(h_0)}\,.
\label{f0h0}
\ee
The results of the numerical computations are presented in Fig.~1,
as a plot of $\log [1-\sqrt{t^*(h_0)}]$ versus $\log(h_0)$. 
According to (\ref{1stf0}) we expect the large $\log(h_0)$ 
slope to be  $(-1)$. We see that this is indeed the case.


\newpage
\begin{thebibliography}{99}

\bibitem{brs} M.~Berkooz, M.~Rozali and  N.~Seiberg,
``Matrix Description of M-theory on $T^4$ and $T^5$,'' 
{\it Phys.Lett.} {\bf B408} (1997) 105, hep-th/9704089.



\bibitem{s9705} N.~Seiberg, "Matrix Description of M-theory on $T^5$  
and $T^5/Z_2$," {\it Phys.Lett.} {\bf B408} 
(1997) 98, hep-th/9705221.


\bibitem{a9911} O.~Aharony, "A brief review of little string 
theories," {\it Class.Quant.Grav.}
{\bf 17}, (2000) 929, hep-th/9911147. 

\bibitem{ms9710} J.~M.~Maldacena and A.~Strominger, 
"Semiclassical decay of near extremal fivebranes,"
{\it JHEP} {\bf 9712}, (1997) 008, hep-th/9710014.

\bibitem{ho00} T.~Harmark and N.~A.~Obers,
"Hagedorn Behaviour of Little String Theory from
String Corrections to NS5-Branes," 
{\it Phys.Lett.} {\bf B485} 
(2000) 285, hep-th/0005021.

\bibitem{br00} M.~Berkooz and M.~Rozali, 
"Near Hagedorn Dynamics of NS Fivebranes, or A New
Universality Class of Coiled Strings,"
{\it JHEP} {\bf 0005}, (2000) 040, hep-th/0005047.

\bibitem{ks00} D.~Kutasov and D.~A.~Sahakyan,
"Comments on the Thermodynamics of Little String
Theory," 
{\it JHEP} {\bf 0102}, (2001) 021, hep-th/0012258.

\bibitem{r0104} M.~Rangamani, 
"Little String Thermodynamics," 
{\it JHEP} {\bf 0106} (2001) 042,
hep-th/0104125. 

\bibitem{gm09} S.~S.~Gubser and  I.~Mitra, 
"Instability of charged black holes in anti-de Sitter space,"
hep-th/0009126.

\bibitem{gm11} S.~S.~Gubser and  I.~Mitra, 
"The evolution of unstable black holes in anti-de Sitter
space," hep-th/0011127.

\bibitem{gl9301} R.~Gregory and  R.~Laflamme, 
"Black Strings and p-Branes are Unstable,"
{\it 
Phys.Rev.Lett.} {\bf 70} (1993) 2837, hep-th/9301052.


\bibitem{gl9404} R.~Gregory and  R.~Laflamme,
"The Instability of Charged Black Strings and p-Branes,"
{\it Nucl.Phys.} {\bf  B428} (1994) 399, hep-th/9404071.


\bibitem{re0104} H.~S.~Reall, 
"Classical and Thermodynamic Stability of Black Branes," 
hep-th/0104071.

\bibitem{chv1} Ali H.~Chamseddine and  Mikhail S.~Volkov,
"Non-Abelian BPS Monopoles in N=4 Gauged Supergravity,"
{\it Phys.\ Rev.\ Lett.} {\bf 79} (1997) 3343, hep-th/9707176.

\bibitem{chv2} Ali H.~Chamseddine and  Mikhail S.~Volkov,
"Non-Abelian Solitons in N=4 Gauged Supergravity and Leading Order
String Theory,"
{\it Phys.\ Rev.\ } {\bf D57} (1998) 6242, hep-th/9711181.


\bibitem{mn0008} J.~Maldacena and C.~Nunez, "Towards the large 
N limit of pure N=1 super Yang Mills," {\it Phys.\ Rev.\ Lett.\ } {\bf 86} 
(2001) 588, hep-th/0008001.


\bibitem{bf0103} A.~Buchel and  A.~Frey, "Comments on 
supergravity dual of pure N=1 Super
Yang Mills theory with unbroken chiral symmetry," hep-th/0103022.


\end{thebibliography}
\end{document}